\documentstyle[aps,draft,epsf]{revtex}

\begin{document}

%
%

\title{ The Prediction and Detection of UHE Neutrino Bursts.}
\vspace{1.3cm}
\author{Mou Roy, H.J. Crawford and Athena Trattner\\ }
\address{ MS 50-245, Lawrence Berkeley National Laboratory,\\
         1 Cyclotron Road, Berkeley, CA 94720.}
\date{\today}
\baselineskip 21 pt

\maketitle

\begin{abstract}

\noindent
In this paper we look at $ \sim 10^{14}$ eV 
neutrino bursts which are predicted to be
observed in correlation with gamma ray bursts (GRBs). 
We describe an efficient way of
constructing a ${\rm km}^2$ effective area detector for these neutrino bursts.
The proposed detector will
cost $<$ \$3M and will operate in
the 4-km deep water off St. Croix for at least one year, sufficient time to
collect the expected $\sim$ 20 events in coincidence with satellite measured 
GRBs provided the fireball model is correct. Coincident gamma ray and neutrino
bursts can be used to test the limits of the relativity principles.

\end{abstract}

\section{Introduction}

Gamma ray bursts are presently the most enigmatic astrophysical
phenomenon. Recent observations indicate that they originate from
cosmological sources \cite{kevin}. 
They are observed in satellites near the Earth at a rate of $\sim$ 1
per day. The relativistic fireball model is consistent with all
observed features of GRBs and has been used by Waxman and Bahcall \cite{bahcall} to predict a measurable flux of $\sim 10^{14}$ eV neutrinos. According to this
model, a detector of $\sim$ 1 ${\rm km}^2$ effective area will observe $\sim$
20 neutrino induced muons per year in coincidence with GRBs.
Other models of
astrophysical processes also demand production of high-energy neutrinos,
including other burst models, AGN models, and topological string models. 
The Neutrino Burster Experiment (NuBE) will measure the flux of UHE ($>$ 1 TeV)
neutrinos over a $\sim$ ${\rm km}^2$ effective area and will
test the fireball model stringently
and uniquely, with an inexpensive, quick and robust
experiment.

\section{UHE Neutrinos Coincident with GRBs, Different Models}

\subsection{Ultra-relativistic Fireball Model }

General phenomenological considerations indicate that GRBs could be
produced by the dissipation of the kinetic energy of a relativistic expanding
fireball.
According to Waxman and Bahcall \cite{bahcall},
a natural consequence of the dissipative cosmological
fireball model of gamma ray
bursters is the conversion of a significant fraction of fireball energy
into an accompanying burst of $ \sim 10^{14} $
eV neutrinos, created by photomeson production of pions in interactions
between the fireball $\gamma$ rays and accelerated protons.
The basic picture is that of a compact source producing a relativistic wind.
The variability of the source output results in fluctuations of the
wind bulk Lorentz factor which leads to internal shocks in the
ejecta. Both protons and electrons are accelerated at the shock
and $\gamma$ rays are radiated by synchrotron and inverse Compton
radiation of shock accelerated electrons.
The accelerated protons undergo photomeson interactions and produce
a burst of neutrinos to accompany the GRB.

\setbox4=\vbox to 160 pt {\epsfysize= 5 truein\epsfbox[0 -200 612 592]{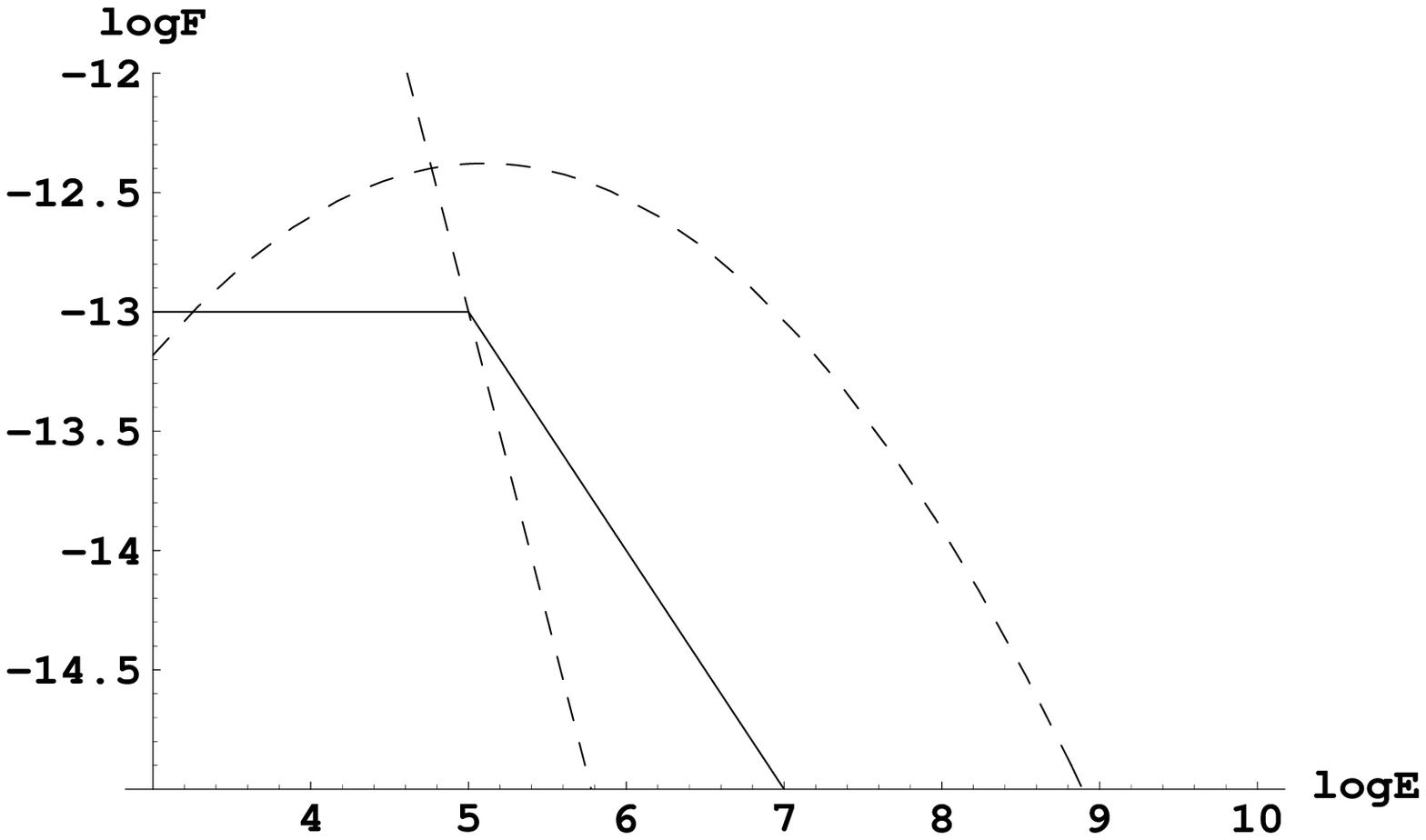}}
\begin{figure}
\centerline{\box4}
\caption{Plot of the expected $ \nu_\mu + {\bar \nu_\mu}$ flux considering
atmospheric background, an Active Galactic Nuclei (AGN) model and the
relativistic fireball GRB model (solid line).
Here F is in ${\rm cm}^{-2} \;\; {\rm s}^{-1} \; {\rm sr}^{-1}$ and neutrino
energy  E has units of GeV.
\label{fig:f.1}}
\end{figure}

\noindent
Figure 1 illustrates the expected neutrino flux from
the GRB model described above in comparison to a typical AGN model
and the expected atmospheric neutrino background.
These neutrino bursts should be easily detected above the
background, since they would be correlated both in
time and angle to the GRB $\gamma$ rays.

\subsection{ Cosmic String Model of Neutrino Production}

UHE neutrinos may also originate in cosmic strings.
Cosmic strings are topological relics from the early universe which
could be superconducting and carry electric current under certain circumstances.
A free string (a nonconducting string uncoupled from electromagnetic
and gravitational fields) generically attains the velocity of light
at isolated points in time and space, which are known as cusps.
Superconducting cosmic
strings (SCS) emit
energy in the form of classical electromagnetic radiation and ultra-heavy
fermions or bosons which decay or cascade at or near the cusp.
Using recent progress on the nature of electromagnetic symmetry
restoration in strong magnetic fields, the study of the decay
products of ultraheavy
fermions near SCS cusps consistent with an SCS explanation of $\gamma$
ray bursts shows that the energy emitted from the cusps is found to be mostly
in the form of high energy neutrinos
\cite{plaga}. The neutrino flux
is roughly nine orders of magnitude higher than that of the $\gamma$ rays.
Therefore this is another model that predicts high energy neutrinos to be
observed in coincidence with $\gamma$ ray bursts.

\section{Detection of UHE Neutrino Bursts}

\subsection{Description of the Detector}

The neutrino burst experiment (NuBE) is designed to search for
UHE neutrinos in coincidence with GRBs.
NuBE is a water Cherenkov detector whose simple design derives from the
very high energy of the GRB neutrinos.
The expected energy of the neutrinos in the fireball model
is $\sim$ 100 TeV, which leads to Cherenkov signals detectable with high
efficiency at large distances from the core track.
We can detect a  muon core when the muon
neutrino interacts in material within $\sim$ 10 km of the array,
leading to a highly radiative muon observable with high efficiency
at perpendicular distances $>$ 150 m from the core all along its multi-km
length. An electron neutrino has a core track which is itself only a few meters in length, but the light from this short core is intense and can be seen
by the proposed array at distances in excess of 500 m.
Coincidence is required with measurements of the photon arrival time and the
GRB location provided by detection in satellites.

\vspace{2cm}

\setbox4=\vbox to 250 pt {\epsfysize= 5 truein\epsfbox[0 -200 612 592]{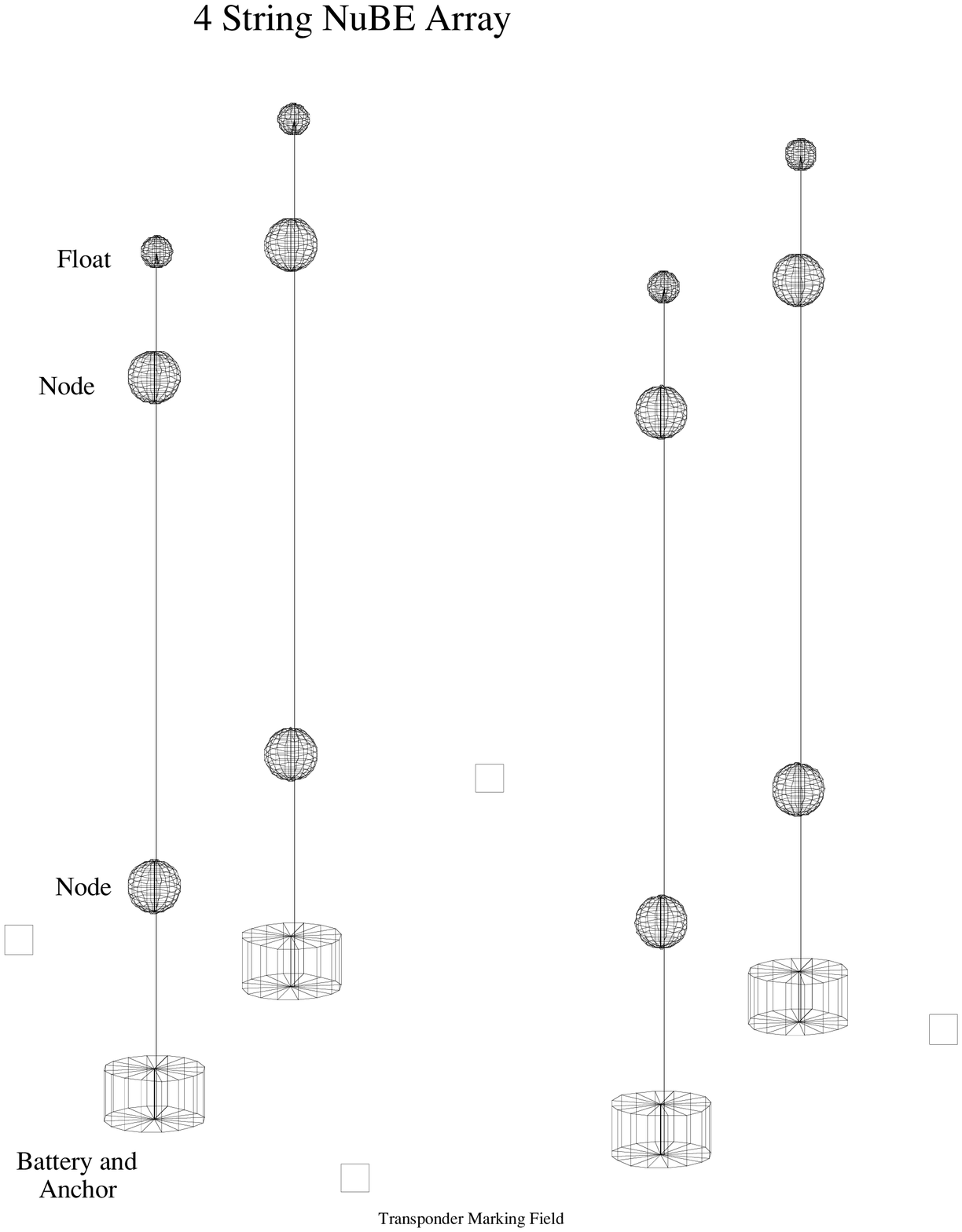}}
\begin{figure}
\centerline{\box4}
\caption{Schematic Diagram of the 4 String NuBE Detector.
\label{fig:f.2}}
\end{figure}

The 4$\pi$ NuBE detector approximates a sphere of diameter $>$ 700m, 
creating an 
effective area of $>$1.5 ${\rm km}^2$. The detector consists of four 
strings placed in the clear water of the deep ocean with their 
anchors at the corners
of a square having $>$ 400m sides, as shown in Figure 2.
Each string has two photon-detector nodes separated by $>$ 400m along the
string. Each node acts independently of the other 7 nodes in the array,
having its own local trigger and data acquisition and storage, thus
providing robustness and redundancy. Local node clocks are periodically
synchronized using bright flashes of blue light from calibration spheres
at each node. Absolute time is kept via these clocks to
accuracy of $<$ 1 second per year. A signal of a high energy event in NuBE
consists of a locally triggered event in
any node occurring within 5 $\mu$s of a locally triggered event in any 
other node. The 5 $\mu$s accounts for the muon or photon transit
time across the array. The coincidence that indicates high energy
events is determined off-line. In a 2-node event the time difference between the
arrival times at each node gives the incident
track direction on a cone, while events having more nodes
hit provide the incident direction to within as little as
$\sim$ 10 degrees. This angular resolution capability provides
robust verification of correlation with the GRB for any signal
that arrives in the GRB time window. The electronics connecting
the photon detector to the data acquisition system is straightforward, 
requiring no further R\&D. It includes a 5ns resolution TDC for relative
arrival time which allows up/down discrimination on local events 
and provides a measurement of the number of photoelectrons produced
by the time over threshold for the signal.

\setbox4=\vbox to 250 pt {\epsfysize= 5 truein\epsfbox[200 -200 612 592]{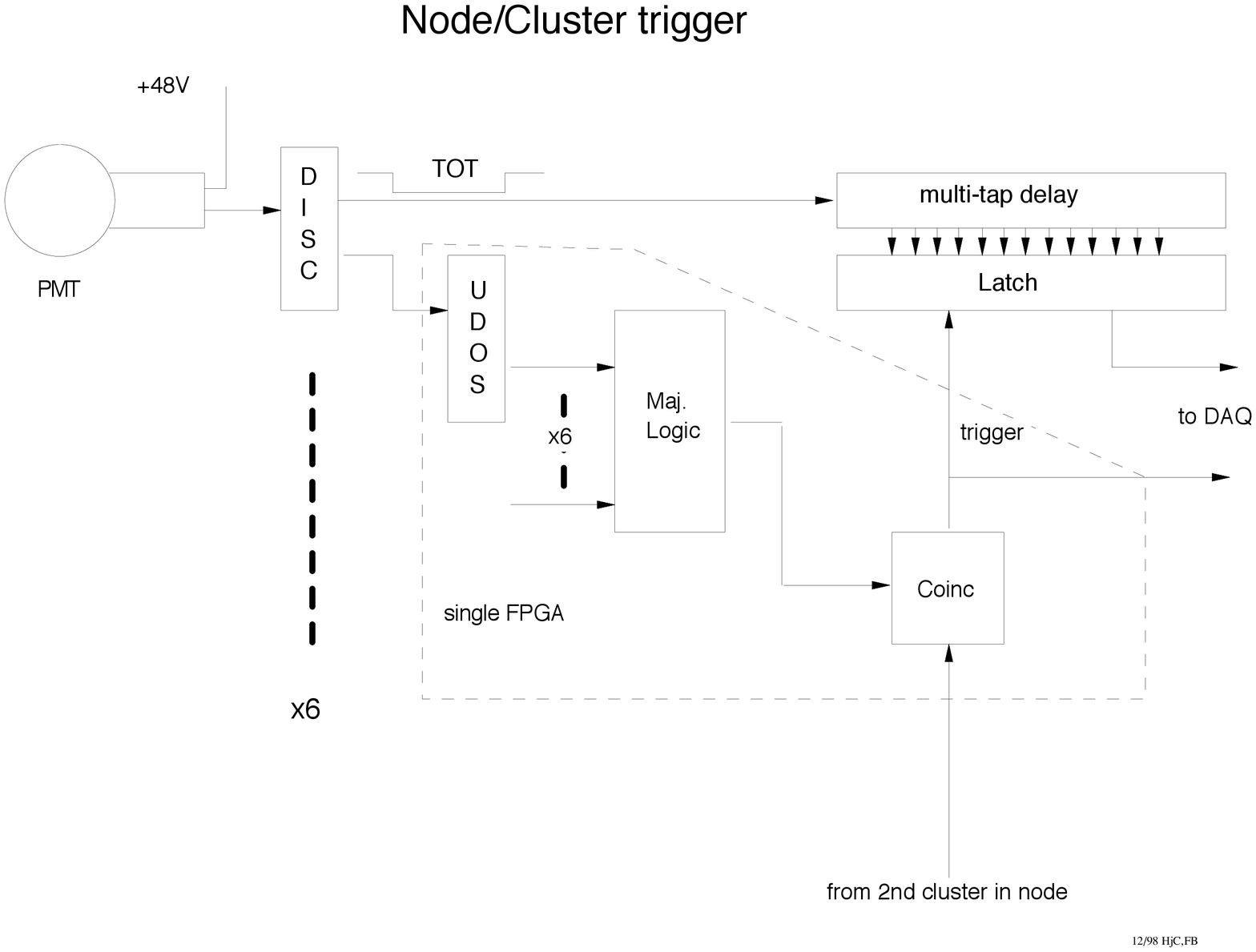}}
\begin{figure}
\centerline{\box4}
\caption{ Systematic Diagram of the NuBE Node/Cluster Trigger 
\label{fig:f.3}}
\end{figure}

Much of the detector can be assembled from ``off-the-shelf" items; anchors,
strings and housing spheres are items of commerce familiar to many of
our collaborators. Deep-sea rated battery packs on each node
can provide $>$ 1 year of untended operation. The detector is easy
to deploy and to recover in any of
a variety of locations, since it doesn't require accurate positioning.
Placing the strings in the location at a site off the coast of St. Croix
in the US Virgin islands can be done easily by vessels of opportunity or with
minimum schedule lead time. This site has the additional advantage of 
providing 4km deep water within 15km of the shore, a clear virtue for site 
visits and data retrieval.
The 4km depth attenuates the cosmic ray muon background to a few per minute
per node, an ideal calibration rate. Up/down discrimination allows us
to calibrate with the Superkamiokande experiment for their highest energy upward
signals, giving verification of an energy threshold for each node.
NuBE provides $>$ 1 ${\rm km}^2$ collecting area in its 4-string implementation and can tell us quickly whether the fireball model is correct in its 
predictions of high-energy neutrino bursts. The total project, from
initial approval to completion of data analysis, will take $<$ 3 years and cost
$<$ \$ 3M.

\subsection{Underlying Physics in the Detection of Neutrino Bursts}

It is important to maximize the physical inferences that can be drawn from
coincident photon and neutrino detection \cite{weiler}. 
The possible simultaneous or near simultaneous observation of neutrinos and
$\gamma$ rays will provide many important new insights into the properties
of neutrinos and GRB sources. It will also yield a novel test of the
weak equivalence principle (WEP).
This has been previously noted for the SN1987A explosion where neutrinos
were observed within a known (but large) time interval of $\gamma$ ray
emission. The same tests can be done for a much higher accuracy and
with better statistics since we will be dealing with multiple
sources at cosmological distances.
The fact that the mystery of the
distance scale for GRBs has been solved in some sources makes this
statement stronger.

The neutrinos from GRBs can be used to test the limits of the relativity
principles. This was done for the neutrino emission from
SN1987A \cite{snequi}.
Neutrinos from GRBs could be used to test the simultaneity of neutrino 
and photon arrival to an accuracy of 1s (1ms for short bursts), 
checking the assumption
that photons and neutrinos should have the same limiting speed. 
Considering a burst at $\sim$ 100 Mpc with $\sim$ 1s accuracy, as an example,
the fractional difference in limiting speed of $\sim 10^{-16}$ is
revealed. 
This may be compared to the SN1987A value of $10^{-8}$.
According to the WEP, the photons and 
neutrinos should suffer the
same time delay as they pass through a gravitational potential.
If the most influential
gravitational potential along the path is the local galaxy, we can compute
a time difference that would result from various trajectories with respect
to the galactic nucleus, the suspected site of the black hole.
NuBE detection of GRB neutrinos would allow a test of the WEP
to an accuracy of $10^{-7}$. Results from measurements on low
energy neutrinos from the supernova 1987A probed this value to $10^{-2}$
\cite{snequi}. On the other hand the most influential gravitational potential
sampled may be near the source itself. If we see nearly the same delay for all
GRB events regardless of distance this may point to a failure of general
relativity in predicting the exit time from the source.
Since there are several GRB sources the corresponding statistics would increase,
and the GRB sources being much further away than the SN1987A would offer a new
distance scale and improved sensitivity.

It would be interesting to investigate the possible detection of
Tau neutrinos.
This would imply neutrino oscillations in transit because
none of our astrophysical models predict Tau neutrinos to be produced at the
source. The key signature is the charged current Tau neutrino interaction, 
which produces a double cascade, one on either end of a minimum ionising
track. Tau neutrinos could be theoretically identified by the double bang
events \cite{pakvasa} but the two individual bangs would be very difficult
to resolve in our proposed detector.

This experiment could be an indirect test for pointing to the model for
the highest energy cosmic ray production. There have been suggestions that
GRBs could be the source of Ultra High Energy (UHE) Cosmic Rays (CR) 
\cite{waxman1}. This source model is consistent with the observed 
CR flux above $10^{20}$ eV. For a homogeneous GRB distribution this model
predicts an isotropic, time dependent CR flux. 

Thus the large distances, short emission time, and trajectory through varying
gravitational fields, leads to the potential for tests of some fundamental
neutrino properties not possible in terrestrial laboratories \cite{weiler}.
Limits may also be placed on neutrino mass, lifetime, electric charge and
on neutrino oscillation parameters.

\section{Conclusions and Present Status of Field}

There are a number of active efforts to observe high energy astrophysical
neutrinos including AMANDA, NESTOR, Baikal, ANTARES and Superkamiokande. 
The relatively dense instrumentation of these experiments, compared to NuBE,
is intended to 
derive source origin by pointing back to the neutrino trajectory
with a high degree of accuracy. 
The Superkamiokande detector in particular provides an excellent
calibration point for a large area detector because of its very high
efficiency for GeV neutrinos. 
However these are all relatively small arrays and consequently will detect
at best only one or two neutrinos coincident with GRBs per year.
NuBE in comparison is aimed at making a large ($> 1 {\rm km}^2$
effective area ), sparse detector to look specifically for neutrinos $>$ 10 TeV
and to determine whether they are in coincidence
with the GRBs.
It is the most efficient and robust way of constructing a detector
for UHE neutrino bursts and will  detect $\sim$ 20 events per year (as predicted by the
fireball model).


\end{document}